\documentclass[a4paper,twoside]{article}

\usepackage{epsfig}
\usepackage{subcaption}
\usepackage{calc}
\usepackage{amssymb}
\usepackage{amstext}
\usepackage{amsmath}
\usepackage{amsthm}
\usepackage{multicol}
\usepackage{pslatex}
\usepackage{apalike}
\usepackage{enumitem}
\usepackage{url}
\usepackage{balance}
\usepackage[ruled,vlined]{algorithm2e}
\usepackage{SCITEPRESS}     % Please add other packages that you may need BEFORE the SCITEPRESS.sty package.

\begin{document}

\title{Using Learned Indexes to Improve Time Series Indexing Performance on Embedded Sensor Devices}

%\author{}
%\affialiation{}
\author{
    \authorname{David Ding, Ivan Carvalho, and Ramon Lawrence} 
    \affiliation{University of British Columbia, Kelowna, BC, Canada}
}
%\orcidAuthor{0000-0002-6779-4461}}


\abstract{Efficiently querying data on embedded sensor and IoT devices is challenging given the very limited memory and CPU resources. With the increasing volumes of collected data, it is critical to process, filter, and manipulate data on the edge devices where it is collected to improve efficiency and reduce network transmissions. Existing embedded index structures do not adapt to the data distribution and characteristics. This paper demonstrates how applying learned indexes that develop space efficient summaries of the data can dramatically improve the query performance and predictability. Learned indexes based on linear approximations can reduce the query I/O by 50 to 90\% and improve query throughput by a factor of 2 to 5, while only requiring a few kilobytes of RAM. Experimental results on a variety of time series data sets demonstrate the advantages of learned indexes that considerably improve over the state-of-the-art index algorithms.}

\onecolumn \maketitle \normalsize \setcounter{footnote}{0} 

\section{\uppercase{Introduction}}
\label{sec:introduction}

\noindent Embedded systems and IoT devices collect massive amounts of data for use in environmental, industrial, and monitoring applications. This time series data is typically sent over the network to cloud servers for processing. There is increasing focus on processing data on the edge devices where it is collected to reduce network transmissions and power consumption. 

There are multiple systems for indexing time series data on cloud servers \cite{apacheiotdb,yang19}. These systems cannot be adapted to the low-memory and CPU processing present on embedded sensor devices. Specialized embedded indexes such as Microhash \cite{microhash}, and sequential binary index for time series (SBITS) \cite{sbits} demonstrate good index performance with minimal memory usage. SBITS allows for both querying by timestamp and data value. Querying by timestamp is done using a linear interpolation search that outperforms binary search used in prior embedded index methods.

Learned indexing is a technique for computing models of the data distribution to efficiently predict record locations. Various techniques have been used including RadixSpline \cite{Kipf2020} and the Piece-wise Geometric Models (PGM) \cite{Ferragina2020}. These indexes outperform traditional tree-based indexes for server applications. There has been no prior work on adapting learned indexes for embedded indexing.

The contribution of this work is the adaptation of learned indexes for use with time series data collected and processed by sensor devices. Learned indexing techniques using RadixSpline and Piece-wise Geometric Models are implemented and evaluated on actual sensor devices with real-world data sets. Experimental results show a dramatic improvement of time series query performance by a factor of two to five times compared to binary search and linear interpolation used by SBITS. With a small amount of index memory, the query performance is significantly improved demonstrating that learned indexes are valuable for the sensor domain.

\section{\uppercase{Background}}

\noindent Indexing on flash memory has unique characteristics, and many index algorithms have been developed to optimize performance \cite{indexsurvey}. Indexing on sensor devices is typically on flash memory and requires additional optimizations beyond server-based indexing. Data structures must adapt to the unique performance characteristics of flash memory such as different read and write times and the erase-before-write constraint (i.e. no overwriting or in-place writes). Common optimizations minimize writes and use sequential rather than random writes.

A key consideration for sensor indexing is the limited resources. Devices have between 4 KB and 32 KB of SRAM memory, a processor running between 16 and 128 MHz, and flash storage consisting of a SD card or raw flash memory chips. Index structures must minimize SRAM usage to be practical.

Data processing systems for embedded devices such as Antelope \cite{dbeverysensor} and LittleD \cite{littled} allow on-device data processing. Improved query performance is achievable by indexing the time series data. Antelope \cite{dbeverysensor} supports an inline index sorted by timestamp. Record lookup by timestamp is performed using binary search. MicroHash \cite{microhash} also stores data as a sorted time series data file and utilizes binary search for timestamp queries. MicroHash uses range partitioning to also index data values to support queries by data value. PBFilter \cite{pbfilter} uses Bloom filters to summarize data for indexing. By sequentially writing the data and index files, PBFilter outperforms tree and hashing techniques.

SBITS \cite{sbits} is a sequential indexing structure supporting timestamp and data value based queries. Timestamp queries are efficiently done using linear interpolation. The location of a record is predicted based on its timestamp and a linear approximation of the data records stored. Given that sensor data sets are often collected at consistent intervals, location predication using linear interpolation is highly effective and outperforms binary search. Since data pages are ordered by timestamp, any timestamp can be retrieved in $\mathcal{O}(1)$ reads by calculating an offset using the start timestamp, the query timestamp, and the rate at which records are written to data file (based on sampling rate and number of records per block). The linear interpolation algorithm can be extended to support multiple different sensor sampling rates and time periods by storing a linear approximation for each period, however that is not in the current implementation. SBITS also supports efficient querying by data value by using a user-customizable bitmap index. Prior experimentation \cite{embedindexsurvey} demonstrated that SBITS outperforms MicroHash and other embedded indexing techniques.

Given the relative consistency of sensor timeseries data, linear interpolation for location prediction is effective. However, there are cases in real-world data sets where the data is not as regular and cannot be easily approximated by a single linear function. Events such as changing the sampling frequency, power failures, or sensor outages may cause the data set to be less than regular and index performance to degrade. It is valuable to consider improved location predication techniques such as offered by learned indexes.

Learned indexes leverage machine learning models to outperform traditional indexes and were initially used for static, in-memory datasets \cite{Kraska2018}. Early work on learned indexes focused on modeling the empirical Cumulative Distribution Function (eCDF). Examples of such indexes include the RMI \cite{Kraska2018}, RadixSpline \cite{Kipf2020}, PGM Index \cite{Ferragina2020}, and PLEX \cite{Stoian2021}.

The core idea of those learned indexes is the connection between the eCDF and an element's position in a sorted array. Let \(P(A \leq x)\) be the proportion of elements smaller than a key \(x\). Then, the position of the key \(x\) in a sorted array with \(N\) elements is \(pos = \lfloor N \cdot P(A \leq x) \rfloor\). Obtaining an exact formula for \(P(A \leq x)\) is challenging, hence the indexes opt to train different kinds of machine learning models to estimate the eCDF.

eCDF based learned indexes are highly effective for static, in-memory use cases. Marcus et al. benchmarked the RMI, the PGM, and the RadixSpline against state-of-the-art traditional indexes on real-world datasets \cite{Marcus2020}. Their experimental results show that the learned indexes always outperform traditional indexes on look-up time and size, losing just on build time.

Although learned indexes are highly effective on read-only workloads, adapting the indexes to handle updates is a challenge. This is particularly relevant for the sensor use case, as sensors are continually collecting and indexing data. The index must handle a continuous stream of appended data.

Learned indexes such as ALEX \cite{Ding2020} and LIPP \cite{Wu2021} support updates. These indexes are tree-based, and use the models to search the tree. Instead of only approximating the eCDF, tree-based learned indexes also focus on using the models to partition the data evenly and achieve a small tree height. The consequences of a smaller tree height are faster lookups and a smaller memory footprint.

Despite the progress of updatable learned indexes, they do not always beat traditional indexes in mixed read-and-write scenarios. Wongkham et al. tested ALEX, LIPP, and an updatable version of the PGM in multiple scenarios in their benchmark \cite{Wongkham_2022}. Among their conclusions is that space efficiency is not a guarantee of updatable learned indexes. The gains in memory consumption are lowered for ALEX and PGM under updates, and the LIPP consumes more memory than traditional indexes. This could potentially rule out applications of learned indexes to sensors, as they are constantly updated and have very limited memory resources.

\section{\uppercase{Adapting Learned Indexes}}

Applying learned indexing structures for sensor time series data requires supporting the continual increase of the data size as new values are collected. Therefore, we need to adapt learned indexes to support append operations while keeping memory consumption low. Further, learned indexes only apply to indexing the sensor data by timestamp as records are stored in a sorted file by timestamp as they are collected. Learned indexes do not directly apply to indexing the data (sensor values) in the record, which are not stored in sorted order. 

We focus on adapting bottom-up, eCDF based learned indexes. Indexes that are built bottom-up such as the RadixSpline do not require all the data in advance, in contrast to top-down modeling approaches such as RMIs that require all data in advance and cannot be easily deployed for the time series use case. We adapt and deploy two structures for indexing: the RadixSpline and the PGM Index.

The indexing approach is based on SBITS \cite{sbits} and proceeds as follows:

\begin{itemize}
    \item Each data record containing collected sensor values is stored in a buffered page in memory until the page is full.
    \item A full data page is written to the sorted data file in sequential order.
    \item After the page is written, an index record is created storing the timestamp of the smallest record on the data page and the data page index.
    \item The index algorithm must maintain its index structure in memory with a bounded size (often less than 1 KB) and may periodically flush index pages to storage for persistence.
\end{itemize}

Querying by timestamp is performed by using the index. Querying by data value is unchanged from the previous SBITS implementation as learned indexes do not apply to the unsorted data values.

% TODO: Image showing indexing data pages and retrieval

\subsection{Piece-wise Geometric Models}

The Piece-wise Geometric Model index (PGM) is a learned index that approximates the CDF via piecewise linear approximations (PLA) \cite{Ferragina2020}. At the heart of the PGM lies a hyperparemeter \( \varepsilon\) that controls the error bounds for the linear approximations.

The original PGM index is built from the bottom-up recursively. At the first level, the PGM builds a PLA over the set of points \(\{ (x_{i}, i) \}_{i=0 \ldots n}\) using the optimal algorithm proposed by \cite{ORourke1981} and rediscovered by \cite{Elmeleegy2009}. For subsequent levels, the PGM applies the same linear approximation strategy using the keys from the previous level \(\{ (k_{j}, j) \}_{j=0 \ldots s}\) as points for the PLA. The recursion halts when there is a level with exactly one line.

We adapt the PGM by creating an append method described in Algorithm \ref{alg:AddPGM}. Notice that the method does not know the number of levels for the PGM in advance and starts by appending a key to the linear approximation at the bottom level and propagating the updates to the upper levels if necessary. Another modification of our implementation is the use of the Swing Filter algorithm also proposed by 
\cite{Elmeleegy2009}. The Swing Filter does not yield an optimal number of lines, but requires only \(\mathcal{O}(1)\) time and memory to process a point. This is a benefit in the sensor use case, as the optimal Slide Filter  requires \(\mathcal{O}(h)\) memory in its worst-case where \(h\) is the number of points in the convex hull, which might not fit in-memory. Even though we adapted the PGM to support appends, its procedure for finding a key remains unchanged from the original PGM (see Algorithm \ref{alg:QueryPGM}).

\begin{algorithm}
\caption{AppendPGMAdd(pgm, \(x\))}\label{alg:AddPGM}

\(L \gets\) pgm.countLevels();

\(K \gets x\);

 \For{\(i \gets 0\) \textbf{to} \(L - 1\)}{
  \(c \gets pgm.levels[i].countPoints()\);
  
  pgm.levels[i].add(K); \tcp{add implements the Swing Filter}
  
  \eIf{pgm.levels[i].countPoints() \(>\) c }{
   \(K \gets\) pgm.levels[i].getLastKey();
   }{
   \Return{};
  }
 }
 
 \tcc{If the algorithm reached this step we need to create a new level}
 
 pgm.levels[L] = newPGMLevel();
 
 firstK \(\gets \) pgm.levels[L-1].getFirstKey();
 
 lastK \(\gets \) pgm.levels[L-1].getLastKey();
 
 pgm.levels[L].add(firstK);
 
 pgm.levels[L].add(lastK);
 
 \Return{};
\end{algorithm}

\begin{algorithm}
\caption{QueryAppendPGM(pgm, \(x\))}\label{alg:QueryPGM}

\If{\(x < \)  pgm.levels[0].getFirstKey()} {

\Return{NotFound};

}

\(L \gets\) pgm.countLevels();

m\( \gets 0\); \tcp{index of the model at the i-th level}

 \For{\(i \gets L-1\) \textbf{to} \(1\)}{

     \(a \gets \) pgm.levels[i].getSlope(m);
 
    \(b \gets \) pgm.levels[i].getIntercept(m);
    
    pos \(\gets \lfloor a \cdot x + b \rfloor\);
    
    lo \( \gets \max \{ 0, pos - \varepsilon - 1 \} \);

    hi  \(\gets \min \{ pgm.levels[i].countKeys() - 1, pos + \varepsilon + 1 \} \);
    
    m \(\gets \) smallest value of \(j\) such that \( x \geq\) pgm.levels[i-1].getKey(j) and \(lo \leq j \leq hi\);

 }

     \(a \gets \) pgm.levels[0].getSlope(m);
 
    \(b \gets \) pgm.levels[0].getIntercept(m);
    
    pos \(\gets \lfloor a \cdot x + b \rfloor\);

lo \( \gets \max \{ 0, pos - \varepsilon - 1 \} \);

hi  \(\gets \min \{ pos + \varepsilon + 1, pgm.levels[0].countKeys() - 1, \} \);
 
 \tcc{Search is an implementation of binary search or linear search over the original array containing the keys}
 
 \Return{Search(x, lo, hi)};
\end{algorithm}

\subsection{RadixSpline}

The RadixSpline is a learned index that approximates the CDF via a linear spline and a radix table storing spline points \cite{Kipf2020}. The two hyperparameters that shape the RadixSpline are \(\varepsilon\), the error bound for the spline approximation, and \(r\), the size of the prefix of the radix table entries.

The RadixSpline builds an error-bounded linear spline over the empirical CDF using the greedy algorithm proposed by \cite{Neumann2008}. After the spline is built, the prefixes of the spline points are inserted in a radix table. The radix table is a flat array of size \(2^{r}\) where each entry in the table maps to a range in the spline. %The RadixSpline algorithm was adapted for the embedded implementation to support appending data points as they are recorded by the sensor device and updating the radix table.

Querying for a point in a RadixSpline is done in three steps. The first step is to look for the prefix of the key being searched in the radix table, and find the corresponding spline point. Given the spline point, calculate a narrow range of size \(2 \varepsilon\) where the key could be using linear interpolation. The last step is to find the key in the range using linear or binary search, which is a quick operation because \(2 \varepsilon\) is constant.

We adapted the RadixSpline to support appends by implementing a streaming version of the GreedySpline proposed in \cite{Neumann2008}, as described in Algorithm \ref{alg:AddRadixSpline}. After appending a point, we check if we can adjust the last spline segment to correctly approximate the point position within \(\varepsilon\). If that is not the case, we create a new spline segment covering the new point and propagate the new spline point to the radix table (see Algorithm \ref{alg:AddRadixTable}). %The radix table is updated when full by pair-wise merging of adjacent entries

\begin{algorithm}
\caption{RadixSplineAdd(rs, \(x\))}\label{alg:AddRadixSpline}

\(c \gets\) rs.spline.countPoints();

rs.spline.add(x); \tcp{Add implements GreedySpline}

\If{rs.spline.countPoints() \(>\) c }{
   \(K \gets\) rs.spline.getLastKey();
   RadixTableInsert(rs, K, c); \tcp{see Algorithm 4}
    \Return{};
   }
\Return{};
\end{algorithm}

\begin{algorithm}
\caption{RadixTableInsert(rs, K, pos)}\label{alg:AddRadixTable}

\(r \gets\) rs.r; \tcp{radix prefix size}
K \(\gets\) K - rs.minKey;

nB \(\gets\) mostSignificantBit(bitShiftRight(K, rs.shiftSize)); \tcp{number of bits required to fit on the table}

\(\Delta \gets\) max(nB - r, 0); \tcp{difference between the available and required number of bits }

\If{ \(\Delta >\) 0 }{
    \tcc{New key triggers table rebuild with new prefix size in order to fit; we merge the old radix entries considering the new bit shift}
    rs.shiftSize \(\gets\) rs.shiftSize + \(\Delta\);
    
    newTable  \(\gets\) allocateTable( \(2^{r}\)  );
    
     \For{\(i \gets 0\) \textbf{to} \(2^{r}\)}{
        j \(\gets\) bitShiftRight(i, \( \Delta \));
        
        newTable[j] \(\gets\) min(rs.table[i], newTable[j]);
     }
     
     \For{\(i \gets 2^{r - \Delta}\) \textbf{to} \(2^{r}\)}{
        newTable[i] \(\gets\) INT\_MAX;
     }
    rs.table \(\gets\) newTable;    
} 

\tcp{Update entry of the radix table with the new pos}

T \(\gets\) bitShiftRight(K, rs.prefixSize);

rs.table[T] \(\gets\) min(rs.table[T], pos);
 
 \Return{};
\end{algorithm}

\section{\uppercase{Experimental Results}}

The experiments evaluated sensor time series indexing for multiple real-world data sets. The sensor hardware platform has a 32-bit Microchip ARM\textsuperscript{\textregistered} Cortex\textsuperscript{\textregistered} M0+ based SAMD21 processor with clock speed of 48 MHz, 256~KB of flash program memory and 32~KB of SRAM. The hardware board has several different memory types including a SD card and serial NOR DataFlash\footnote{\url{https://www.dialog-semiconductor.com/products/memory/dataflash-spi-memory}} which supports in-place page level erase-before-write. This platform is representative of embedded devices with commonly used 32-bit ARM processors. The serial NOR DataFlash was used to test performance on raw memory without a flash translation layer (FTL). Testing on raw memory insures no overhead with FTL maintenance operations and allows for the highest read performance. Platform performance characteristics are in Table \ref{hardware}.

The experiments benchmark four indexes:

\begin{itemize}
\item
  \textbf{Binary Search}: A binary search over the sorted data. This is the baseline for the benchmark and requires no additional memory.
\item
  \textbf{SBITS}: SBITS using linear interpolation requires 8 bytes to maintain linear approximation. SBITS was optimized to default to binary search if its predictions were off significantly. %\footnote{Implementation available at \url{https://github.com/ubco-db/sbits}. TODO make link public.}
 \item
  \textbf{PGM}: A modified version of the PGM Index with support for appends and error-bound of \(\varepsilon=1\). %\footnote{Implementation available at \url{https://github.com/IvanIsCoding/AppendPGM}}
 \item
  \textbf{RadixSpline}: A modified version of the RadixSpline with support for appends, error-bound set to \(\varepsilon=1\) and radix prefix size of $r=0$ that offered the best performance with smallest memory usage. %Although multiple values of \(r\) are tested, increasing \(r\) did not provide better metrics, so we report the value for \(r = 0\) as it is the value that uses no radix table and consumes the least memory. %\footnote{Implementation available at \url{https://github.com/ubco-db/spline}. TODO make link public and update code }

\end{itemize}

\begin{figure}[!htbp]
    \includegraphics[width=\linewidth]{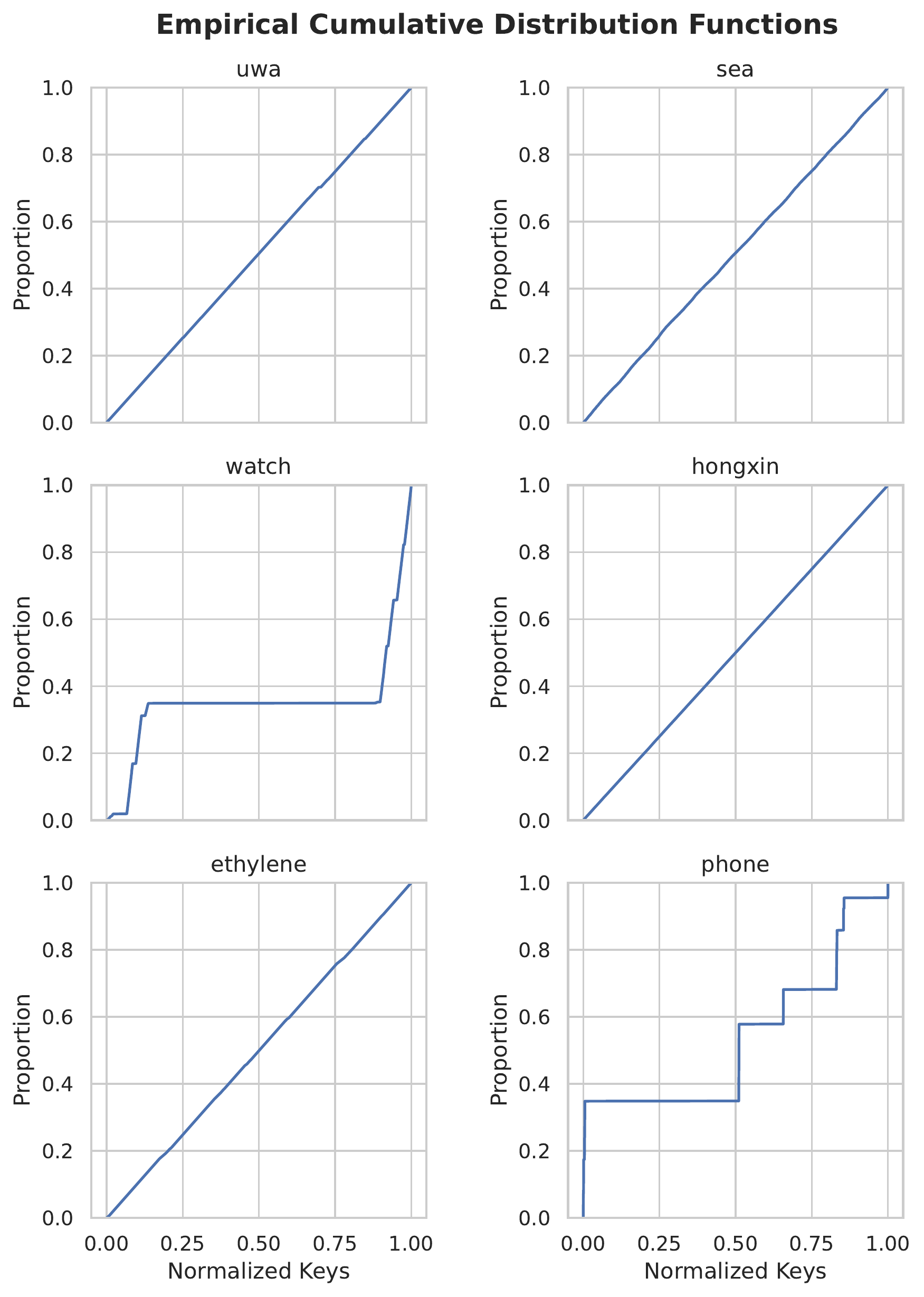}
    \caption{eCDFs for the experimental datasets. The PGM and the RadixSpline use a collection of linear models to approximate those with bounded-error of \(\varepsilon\).}
    \label{fig:ecdf}
\end{figure}

The experiments measure four core metrics: the query throughput, the number of I/Os per query, memory consumption, and insertion throughput. These metrics are relevant to common use cases of sensors such as querying by timestamp and ingesting new data. We report the average of the metrics based on three separate runs. The real-world data sets evaluated are in Table \ref{tbl:data}. The data sets cover a variety of sensor use cases including environmental monitoring, smart watches, GPS phone data, and chemical concentration monitoring. The environmental data sets {\bf sea} and {\bf uwa} were originally from \cite{microhash} and have been used in several further experimental comparisons \cite{sbits,embedindexsurvey}. We also present the eCDF for the datasets in Figure \ref{fig:ecdf}.

\begin{table*}[tb]
\centering
\begin{tabular}{|c|c|c|c|c|c|}
\hline
 & \multicolumn{2}{|c|}{\bf Reads (KB/s)} & \multicolumn{2}{|c|}{\bf Writes (KB/s)} & \\
 \hline
 & {\bf Seq} & {\bf Random} & {\bf Seq} & {\bf Random} & {\bf Write-Read Ratio} \\
\hline
M0+ SAMD21 (DataFlash) & 475 & 475 & 38 & 38  & 12.5 \\
\hline
\end{tabular}
\caption{Hardware Performance Characteristics}
\label{hardware}
\end{table*}

\begin{table*}[tb]
\centering
\begin{tabular}{|c|c|c|c|c|c|}
\hline
{\bf Name} & {\bf Points} & {\bf Points Used} & {\bf Sensor Data} & {\bf Source} \\
\hline
sea	    & 100,000   & 100,000   & temp, humidity, wind, pressure    & SeaTac Airport \\
uwa	    & 500,000	& 500,000   & temp, humidity, wind, pressure    & ATG rooftop, U. of Wash. \\
hongxin	& 35,064    & 35,000    & PM2.5, PM10, temp	                & \cite{Zhang2017} \\
ethylene &	4,085,589   &  100,000  & ethylene concentration        &  \cite{Fonollosa2015} \\
phone   & 18,354    & 18,000    & smartphone X/Y/Z magnetic field	&  \cite{Barsocchi2016} \\
watch	& 2,865,713 & 100,000   & smartwatch X/Y/Z gyroscope                   & \cite{Stisen15}\\
\hline
\end{tabular}
\caption{Experimental Data Sets}
\label{tbl:data}
\end{table*}

\subsection{Query Performance}

One common use case of processing data in edge devices is to query timestamps. We measure the query throughput for searching timestamps. After the sensor data was inserted, 10,000 random timestamp values were queried in the timestamp range. Data was collected on the time to execute all queries and the number of I/O operations performed.

\begin{figure}[!htbp]
    \includegraphics[width=\linewidth]{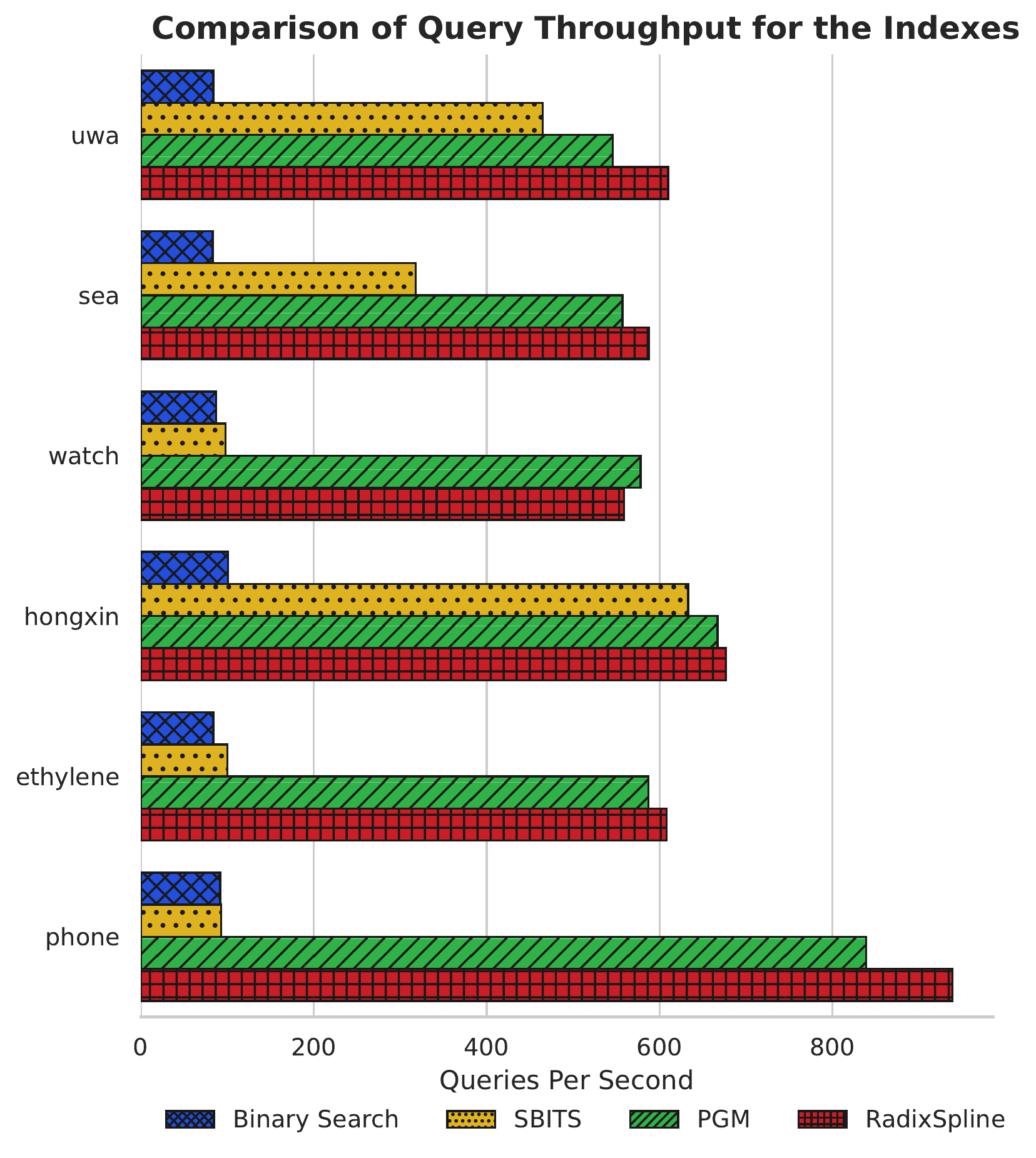}
    \caption{Average query throughput among the indexes for each dataset in the benchmark. Higher rates indicate better results. The RadixSpline and PGM consistently outperform SBITS and binary search.}
    \label{fig:qps}
\end{figure}

The query throughput results are in Figure \ref{fig:qps}. Applying learned indexes to sensor time series is very effective. The PGM and the RadixSpline always outperform SBITS with significant improvements on highly variable datasets.

For the \textbf{hongxin} and \textbf{uwa} datasets, the gains are more modest ranging from 1.05x-1.20x for the PGM and 1.06x-1.31x for the RadixSpline. These datasets are highly linear such that an interpolation search achieves its best-case scenario. Since the PGM and RadixSpline use PLAs to model the dataset, they also perform well just like the interpolation search.

However, datasets such as \textbf{watch}, \textbf{ethylene}, and \textbf{phone} prove to be more challenging for SBITS. Its performance is affected and becomes closer to the throughput of the binary search baseline. Learned indexes, on the other hand, extend their lead and stay resilient to the change in data distribution. The performance gains range from 1.8x-8.9x for the PGM and 1.8x-10x for the RadixSpline. This indicates that the learned indexes have more flexible models that describe the data distribution better.

The query throughput difference between RadixSpline and PGM is within 10\% for all data sets. Since both approaches use linear approximations, the difference relates to how the linear approximations are themselves indexed. For these experiments, the radix table for RadixSpline was allocated no space, and only a binary search was used on the spline points. This is effective as there are very few points. Performance testing with using a radix table of size $r=4$ and $r=8$ demonstrated no query performance benefit while consuming precious RAM. This makes sense as the radix table is only saving a few comparisons when searching the spline points in memory, and the search time is dominated by the I/Os performed to the flash memory. PGM produces a multi-level index, which takes some more space and a little longer to query. Overall, both approaches are effective and greatly improve on binary search or single linear interpolation. The trade-off between the query performance and memory space is discussed further in Section \ref{sect:memory}.

\begin{figure}
    \includegraphics[width=\linewidth]{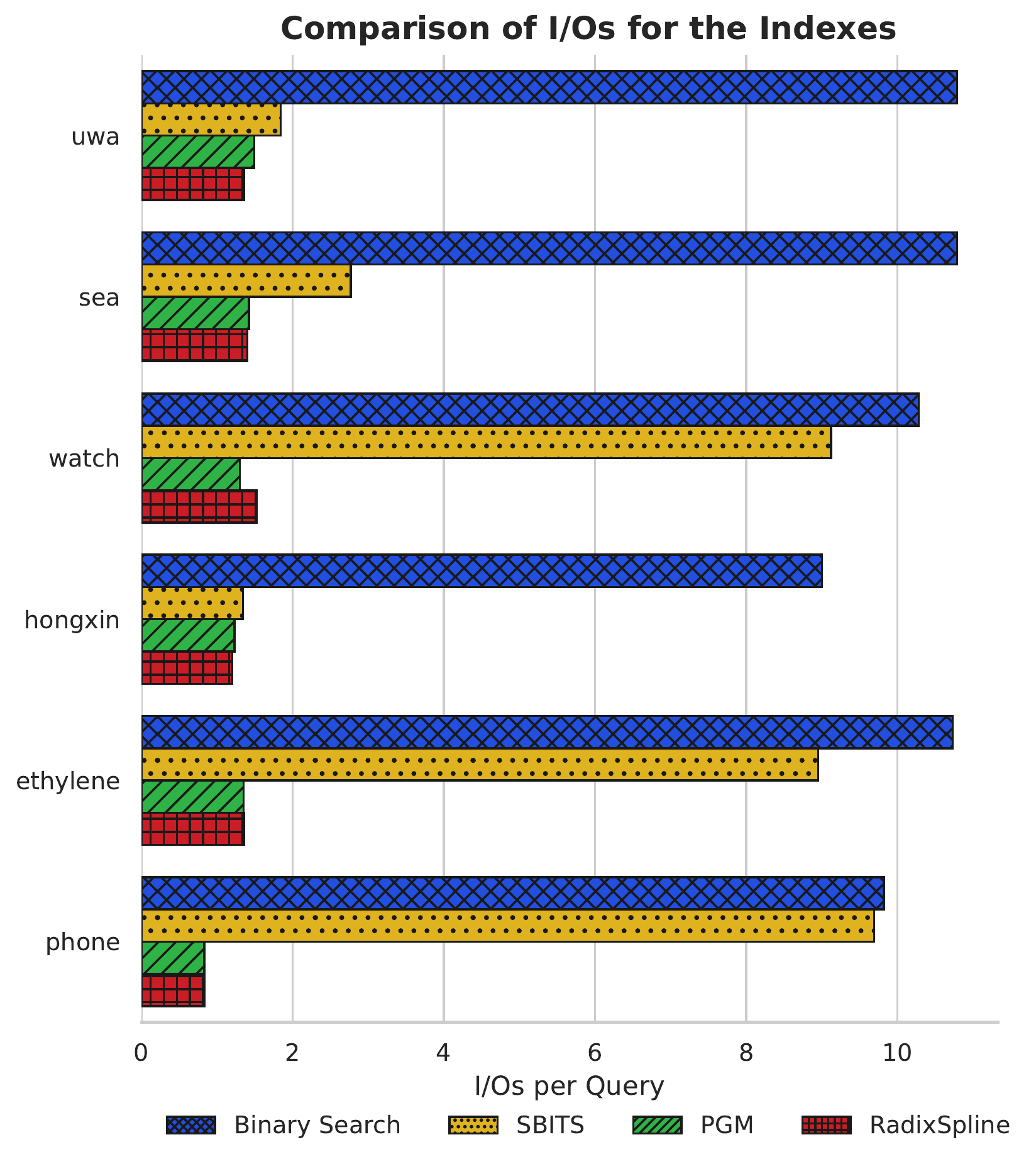}
    \caption{Average number of I/Os per query among the indexes for each dataset in the benchmark. Lower rates are better. Learned indexes significantly reduce the number of required I/Os.}
    \label{fig:ioq}
\end{figure}

Another relevant metric for querying is the number of I/Os per query. The number of I/Os performed dominates the query response time and is especially important in embedded systems were predictable real-time performance is desirable. Figure \ref{fig:ioq} displays the average number of I/Os performed per timestamp query. Binary search performs \(\mathcal{O}(\log N)\) I/Os and has significantly more I/Os than the other approaches. SBITS' single linear interpolation is effective in many cases, however data sets like phone are poorly approximated by one linear approximation and the algorithm frequently defaults to binary search due to poor prediction accuracy. Both PGM and RadixSpline have guaranteed error bounds in their construction. With \(\varepsilon=1\), at most 2 I/O are performed for any lookup with the average often around 1.3 to 1.5. This predictable performance is a major benefit for using these learned indexes.

\subsection{Memory Space Efficiency} \label{sect:memory}

It is critical for learned indexes to have a small memory footprint in order to be useful for embedded systems. Many traditional techniques cannot be applied to embedded systems because of the RAM constraints \cite{sbits}, which prompts adaptations that trade memory and performance. The memory results are in Table \ref{memory_consumption}. 

\begin{table}[tbh!]
\centering
\begin{tabular}{|c|c|c|}
\hline
\multicolumn{3}{|c|}{\bf Memory Consumption (in KB)} \\
\hline
 & {\bf PGM} & {\bf RadixSpline} \\
\hline
uwa & 0.04 & 0.09 \\ \hline
sea & 1.88 & 0.96 \\ \hline
watch & 7.80 & 2.43 \\ \hline
hongxin & 0.10 & 0.07 \\ \hline
ethylene & 0.40 & 0.12 \\ \hline
phone & 1.09 & 0.25 \\
\hline
\end{tabular}
\caption{Memory consumption comparison among the learned indexes for each dataset in the benchmark.}
\label{memory_consumption}
\end{table}

All indexes fit in memory, with the maximum amount of memory used being 7.80 KB. The results indicate that the RadixSpline consumes less memory than the PGM. This is to be expected, as both the spline and the bottom level of the PGM contain a very similar set of linear approximations. The key difference between the two is their approach to finding the linear approximation to use. The RadixSpline uses a small radix table to index spline points. The PGM uses a recursive approach building additional PLAs. Since the size of the table from the RadixSpline is parametrized by \(r\), it is possible to sacrifice a little query performance to achieve less memory usage.

For our choice of \(r = 0\), the difference in memory consumption is significant, and the difference in query performance is negligible as performance is dominated by the number of I/Os not comparisons in memory. The PGM uses two to four times the amount of memory as the RadixSpline for five out of the six datasets, with the exception being the easiest dataset \textbf{uwa}. 

By modifying the error bound ($\epsilon$), both approaches can reduce their memory footprint at the sacrifice of more query I/Os and lower query throughput. Table \ref{error} shows statistics on the index size in bytes, I/Os performed per query, and query throughput in queries/second for the {\bf sea} data set for multiple different values of $\epsilon$. There is a quite significant index size reduction for increasing $\epsilon$ to 2 or 3. Even though the I/Os per query increases, it is always bounded by $\epsilon$. This allows designers to determine the exact performance trade-offs in terms of space and query I/Os in a predictable fashion.

\begin{table*}[tb]
\centering
\begin{tabular}{|c|c|c|c|c|c|c|}
\hline
 & \multicolumn{3}{|c|}{\bf RadixSpline} & \multicolumn{3}{|c|}{\bf PGM} \\
 \hline
 {\bf Test} & {\bf Index Size} & {\bf Query I/O} & {\bf Throughput}  & {\bf Index Size} & {\bf Query I/O} & {\bf Throughput} \\ 
\hline
SEA $\epsilon=1$	& 932   & 1.40  & 588   & 1920  & 1.42  & 557 \\
SEA $\epsilon=2$	& 436	& 1.85	& 461	& 856	& 1.91	& 440 \\ 
SEA $\epsilon=3$	& 260	& 2.23	& 386	& 664	& 2.26	& 376 \\
SEA $\epsilon=5$	& 212	& 3.12	& 281	& 496	& 3.41	& 256 \\
SEA $\epsilon=10$   & 132	& 5.37	& 166	& 232	& 4.94	& 179 \\
\hline
\end{tabular}
\caption{Index Size in bytes, I/Os per Timestamp Query, and Query Throughput (queries/sec.) for Different Error Bounds $\epsilon$}
\label{error}
\end{table*}

\subsection{Insertion Performance}

\begin{figure}
    \includegraphics[width=\linewidth]{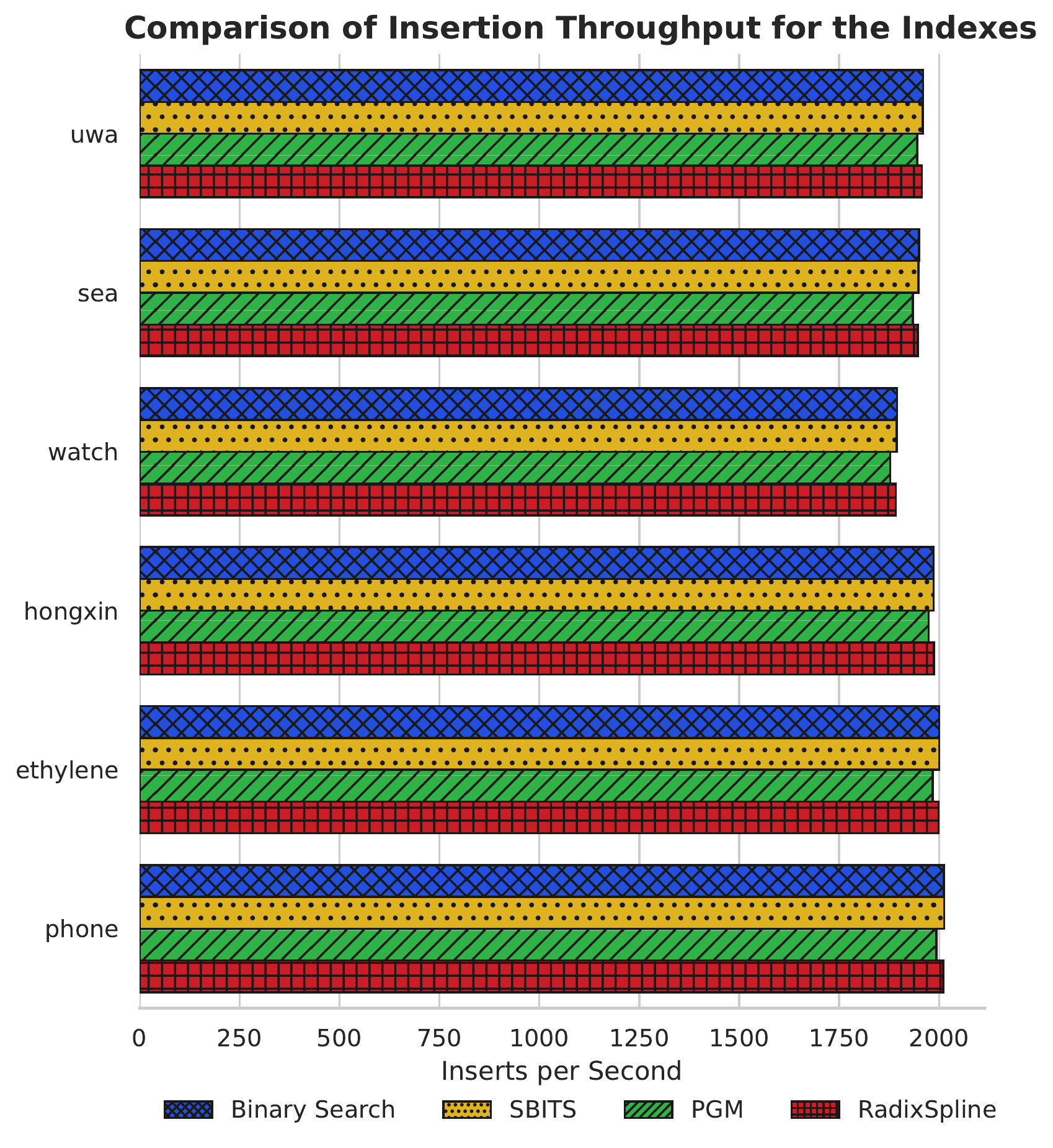}
    \caption{Average insertion throughput among the indexes for each dataset in the benchmark. Higher rates indicate better results. The insertion throughout is slightly lower for the PGM, but overall learned indexes remain competitive.}
    \label{fig:insert}
\end{figure}

Adding indexes benefits query performance, but they can also impact the insertion time. It is critical for sensors to keep ingesting data; hence the need to monitor the insertion throughput to ensure they match the sensors required sampling rate. For insertion performance, the $N$ records used for each data set were inserted at the maximum possible rate of the hardware. The insertion performance is dominated by the I/Os for writing the data pages to storage, but the index construction time may have some overhead. The insertion rates in Figure \ref{fig:insert} show the maximum rates possible on the hardware for each data set and index.

The baseline for insertion performance is the binary search case which consists of a simple data record append and no indexing overhead. The average throughput was 1967 inserts per second. Since the incoming timestamps are strictly increasing and because binary search does not store any additional information to help on the search, this represents the upper bound for insertion performance.

SBITS and RadixSpline were the indexes that had the highest insertion throughputs at 1966 and 1965 inserts per second, respectively. They almost match the scenario with no index at all. The results are consistent with earlier experiments for SBITS \cite{embedindexsurvey}, showing that the number of I/Os to keep the index up to date is minimal. Perhaps the most surprising result is for the RadixSpline, because it needs to calculate spline points and update the radix table. Our benchmark shows that the overhead for these operations are small.

The PGM supports 1951 inserts per second, which represents a less than 1\% overhead. The PGM has a slightly lower insert rate because some insertions trigger changes on multiple levels of the PGM, while RadixSpline triggers at most one change in the spline and radix table. The PGM insertion throughput remains competitive beating traditional embedded indexes such as B-Trees \cite{embedindexsurvey}.

\subsection{Results Discussion}

Overall, the experimental results demonstrate that there are significant advantages to using learned indexes adapted for embedded time series data. The most significant advantage is the predictable and bounded timestamp query performance. By specifying a given error bound (\(\varepsilon\)), the maximum number of I/Os per query is \(2\varepsilon+2\). The query performance is significantly higher than SBITS linear interpolation search or binary search. The overhead of the index in terms of insertion throughput is minimal. To handle the limited memory, the index size can be reduced by increasing the error bound. In the experiments tested, the index size was usually less than a few KB. The index algorithms support real-time sensor data collection of almost 2000 records/second on the experimental hardware, which is significantly above collection rates for most applications.

\section{\uppercase{Conclusions}}

This work adapted and applied learned indexes to sensor time series data. Experiments indicate that the RadixSpline and PGM are suitable to tackle the challenge of indexing the amount of data embedded systems collect under the unique constraints that those systems have. This work is a bridge between embedded system indexing and the original applications of learned indexing targeting large in-memory databases. With the correct adaptations, it is possible to reuse these structures conceived to query large static datasets and apply them in embedded systems with tiny memory and frequent data appends.

Future work will extend learned indexes for sensors to automatically handle hyperparameters and require less tuning. We envision learned indexes that automatically adapt their parameters to fit in the limited resources in embedded systems, specifically adapting the error bound to ensure the index fits in the memory available.

\balance

\bibliographystyle{apalike}
{\small
\bibliography{refs}}

\end{document}